\documentclass[final,3p,times,twocolumn]{elsarticle}
\usepackage{lineno}
\journal{Nuclear Instruments and Methods A}
\usepackage{graphicx}
\usepackage{upgreek}
\usepackage{amsmath, amssymb}
\usepackage{subfigure}          
\usepackage[T1]{fontenc}
\usepackage{textcomp}

\newcommand{\glq}{\textquotesingle}
\newcommand{\grq}{\textquotesingle}
\def\um{\mathrm{\,\upmu m}}

\begin{document}

\begin{frontmatter}

\title{Pixel Detectors ... where do we stand?}

\author[]{N.~Wermes}
\ead{wermes@uni-bonn.de}

\cortext[cor]{Corresponding author}

\address{University of Bonn, Bonn, Germany\\\hfill }

\begin{abstract}
Pixel detectors have been the working horse for high resolution, high rate and radiation particle tracking for
the past 20 years. The field has spun off into imaging applications with equal uniqueness. Now the move is
towards larger integration and fully monolithic devices with to be expected spin-off into imaging again.
Many judices and prejudices that were around at times were overcome and surpassed. This paper attempts to give an account of the developments following a line of early prejudices and later insights.
\end{abstract}


\end{frontmatter}

\section{Introduction}
Pixel detectors have entered particle physics experiments in the early nineties, first with small scale
devices \cite{Heijne:1992zb,Becks:1997ei} and soon after with developments of detectors with \cal{o}(10$^8$) pixels
for the LHC experiments (see e.g.\,\cite{pixel_book}). Despite early approaches with monolithic devices \cite{Snoeys:1992yh,Dierickx1998,MAPS-epi_Turchetta:2001} so-called hybrid pixels in which
pixel sensor and readout chip are separated entities, mated by employing bumping and flip-chipping
technology, have been the technology of choice, in particular for applications in high rate and
radiation environments \cite{mgsnw2018} as at the LHC. Soon after also developments targeting imaging applications, in particular biomedical and synchrotron light X-ray imaging \cite{Campbell:1997cu,Fischer:1997eu} have been started, commencing their own development branches. As of today these development have culminated in very successful large scale detectors in the LHC experiments \cite{ATLAS-pixel-paper_2008,CMS-pixel-paper_2008,Chochula:2003je,pixel_book}
on the one hand and in imaging detectors on the other \cite{Medipix_recent,Redford:2016evz,Gottlicher:2009jua}.

A bit later though, monolithic pixel detectors also reached a state of performance that made them attractive
for particle and heavy ion physics experiments where low material budget plays an overwhelming role due to the
low average momenta of particles emerging from the interactions. Fully monolithic pixel detectors where realised
in the STAR experiment at RHIC \cite{RHIC_Schambach_2015} and a follow-up 10\,m$^2$ detector is in production
for the ALICE Inner Tracker ITS \cite{Abelevetal:2014dna}. Also for the extreme radiation and rate environment
encountered at the HL-LHC so-called depleted monolithic active pixels (DMAPS) have been developed \cite{Peric:2007zz} \cite{Wermes:2016dav} that are able to cope with them (section~\ref{DMAPS}). As was the case for hybrid pixels monolithic imaging devices are also spinning off, targeting X-ray and synchrotron radiation applications in astrophysics, biomedical imaging, and spectroscopy. This conference has devoted an integrated
workshop on SOI-pixels in particular addressing such applications \cite{Arai_HSTD2017}.

\paragraph{Early prejudices}
In order to put the developments into some perspective I have chosen to follow a personal line of some
early thinking and prejudices on pixel detectors, in particular those developed for LHC rate and radiation levels.
Such statements are for example:
\begin{itemize}
\item[-] Radiation levels at LHC and HL-LHC are tough for pixel sensors; and there is no alternative to planar pixels.
\item[-] Diamond will never become a material suited for a pixel tracker.
\item[-] The bulk material choice has to be p-type.
\item[-] Pixel sensors from CMOS IC fabrication lines do not have sufficient performance in terms of charge collection and radiation tolerance.
\item[-] A complex chip with {\cal{o}}(10$^9$) transistors can only be done by industry, needs many years of development, and is too expensive. However, as the 250\,nm technology was already radhard, the 65\,nm technology will be even better.
\item[-] Only hybrid pixel designs can cope with LHC conditions. Monolithic pixels will never stand the LHC rate and radiation environment.
\item[-] SOI pixel technology is fine, but it is extremely difficult to get around the many challenges of shielding
sensitive structures.
\item[-] Silicon micropattern detectors are good for spatial resolution, but not for timing.
\end{itemize}
Much of this early thinking was wrong or has been overcome today as will be addressed in this paper.

\section{Pixel sensors in high radiation environments}\label{sec:radiation}
\paragraph{\bf N- or p-type bulk}
Apart from rate capability the required radiation tolerance is the dominant
issue to cope with in most high energy physics experiments today.
At the HL-LHC every Si lattice cell will see about 50 particles during its lifetime.
The recipe developed over the years has been (i) readout at n$^+$ electrodes (e$^-$ collection), (ii) operation
at high bias voltages, (iii) a careful planning of the annealing scenario, (iv) proper electrode and
guard ring design, and recently (v) to use p-substrates with n electrodes (rather than n-in-n). While there
is empirical evidence for the latter \cite{Moll:2009zz} it became evident that details of the electrode structure
and sensor thickness complicated the issue \cite{Moll:2009zz}. Only recently p-type silicon as a sensor material
has been studied in some detail. Bombardment with hadrons (p,n,$\pi$) causes damage to the silicon lattice which is independent of the type of doping. However, the effective doping concentration changes due to different effects. While in n-type silicon shallow donors (like phosphorus) are removed when a V-P complex is formed, the acceptor (boron B) abundance seems to be dominantly decreased by an interstitial B$_i$O$_i$ complex \cite{Junkes:2016ztr,Donegani_2017}, decreasing the negative space charge of the bulk. For n-type material a practical and efficient mitigation has been invoked by oxygen enrichment having VO complexes competing with VP. For p-type material such an efficient cure still has to be established \cite{Moll-radecs}.

\begin{figure}
\centering
   \subfigure[Planar pixel cross section]{\raisebox{.5cm}
            {\includegraphics[width=0.60\linewidth]{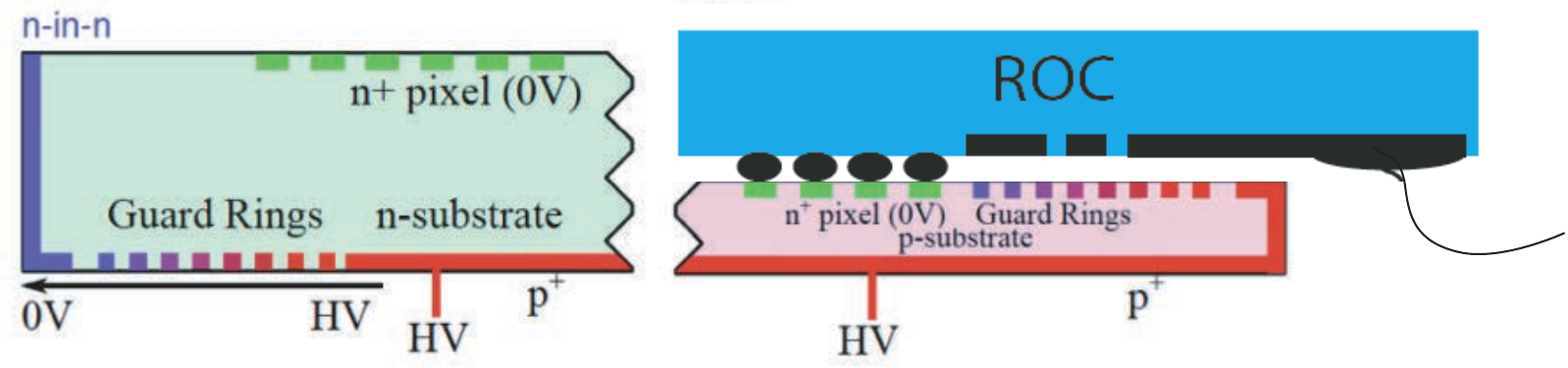}}\label{fig:planar-pixels-a}}
            \hskip 0.0cm
   \subfigure[Radiation hardness]{\raisebox{0.0cm}
            {\includegraphics[width=0.38\textwidth]{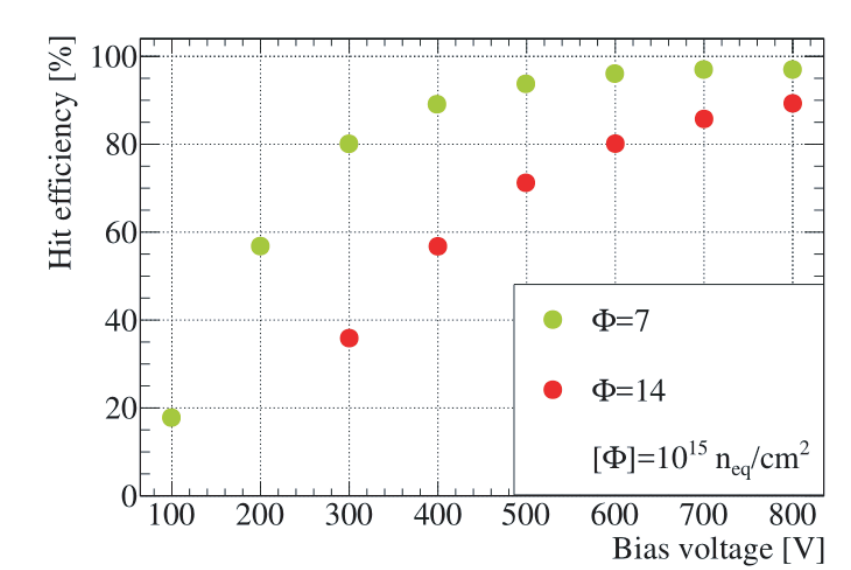}}\label{fig:planar-pixels-b}}
\caption[]{Planar pixel developments. (a) Cross section of planar pixels. (left) current
(250--300\,$\um$) LHC pixels. (right) Thin (100--150$\um$) pixels with readout chip (ROC). (b) Hit efficiency as a function of bias voltage for two very high fluences 7 and 14 $\cdot 10^{15}$n$_{eq}$cm$^{-2}$.}\label{fig:planar-pixels}
\end{figure}

Therefore, while for LHC upgrades the change from $p^+-{\rm in}-n$ to $n^+-{\rm in}-p$ for LHC experiment upgrades
is necessary for strip detectors, from the radiation point of view, $n^+-{\rm in}-n$ (present pixel choice)
is equally suited at least, except that for the fabrication of the latter double sided processing is needed.
Hence strip and hybrid planar pixel sensors for the HL-LHC upgrades both plan on employing p-type substrates. For planar pixel detectors fig.~\ref{fig:planar-pixels-a} shows the cross section developing from the
current $n^+-{\rm in}-n$ LHC pixels (250--300\,$\upmu$m) to thin $n^+-{\rm in}-p$ pixel sensors (100--150$\um$); fig.~\ref{fig:planar-pixels-b} shows their performance under extreme radiation damage. Note in fig.~\ref{fig:planar-pixels-a} that the readout chip (ROC) must be placed in very close distance to parts of the sensor being under high voltage potential. Special isolation protection must be applied.

\begin{figure}
    \centering
    \subfigure[3D-Si design for HL-LHC]{\raisebox{0.3cm}
        {\includegraphics[width=0.58\textwidth]{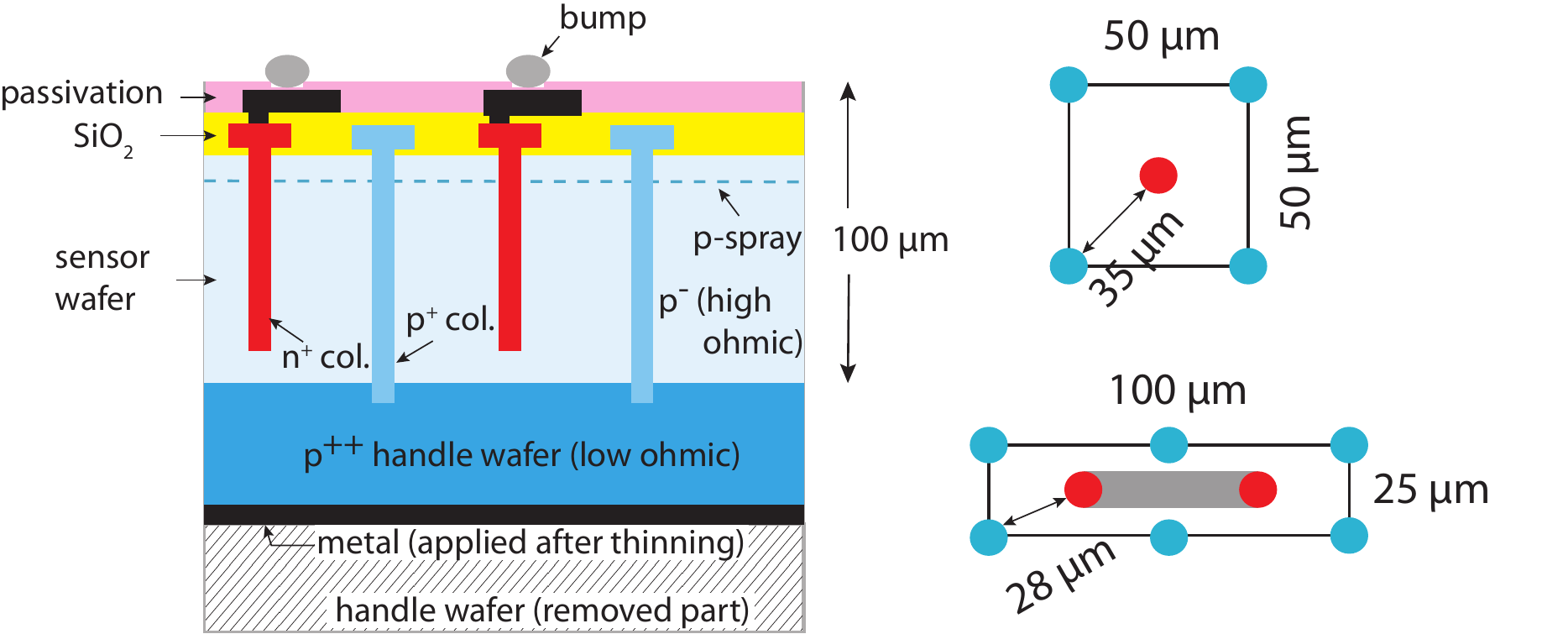}}\label{fig:3D-Si_new}}\hskip 0.0cm
    \subfigure[Performance]{\raisebox{0.0cm}
        {\includegraphics[width=0.40\textwidth]{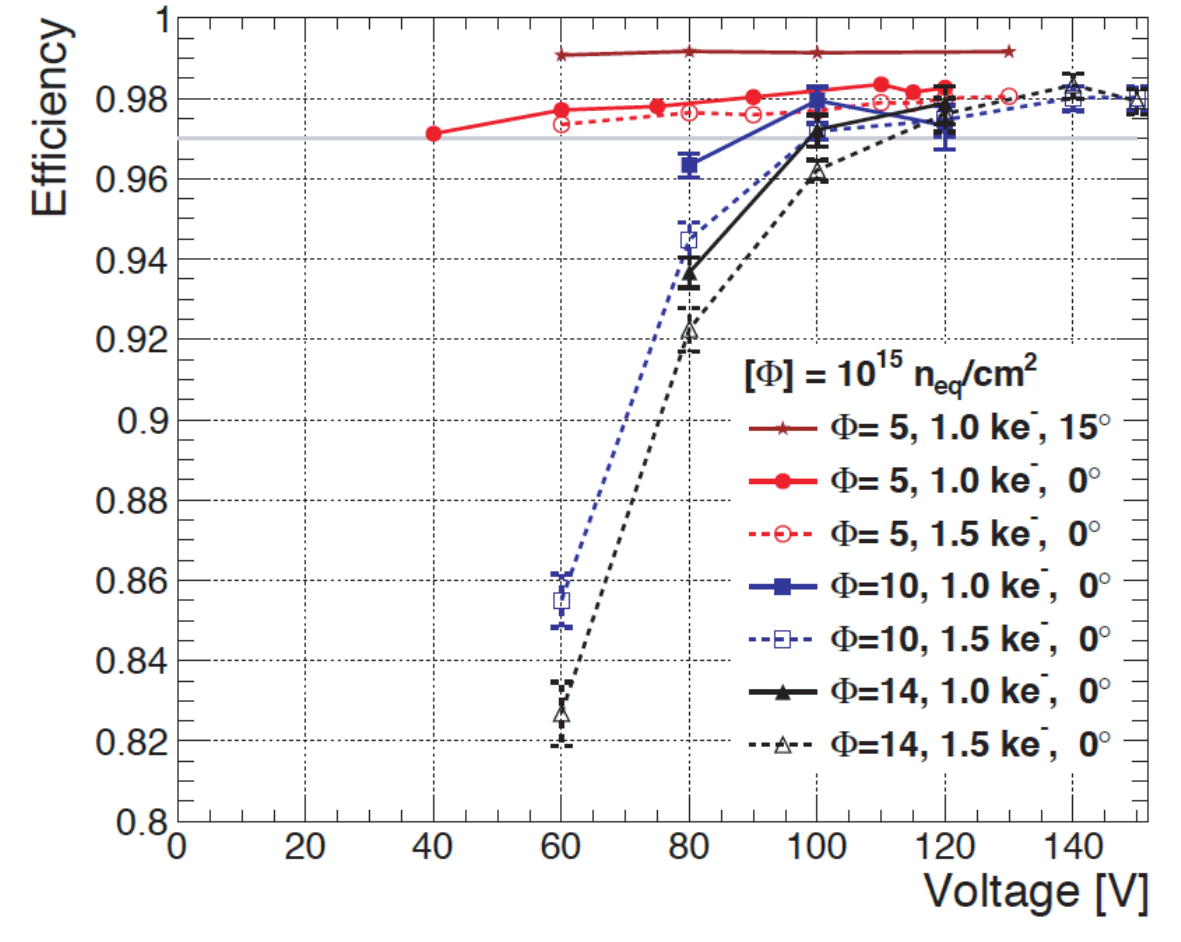}}\label{fig:3D-Si_performance}}
    \caption{3D-Si sensors: (a) Thin 3D-Si design optimized for HL-LHC (adapted from \cite{DallaBetta:2016_review}) with two top view sketches for $50\times 50\,\upmu$m$^2$ and $25\times 100\,\upmu$m$^2$ pixel sizes, respectively \cite{Lange:2016jbm}. While the p$^+$ columns are deep etched through to a handle wafer, the n$^+$ columns stop about 15\,$\upmu$m short. (b) Hit efficiency as a function of bias voltage for different fluences and design variants \cite{Lange:2016jbm}\label{fig:3D-Si_sensors}.}
\end{figure}

\paragraph{\bf 3D pixel electrodes}
Very high fluences have imposed a real threat to standard pixel detector designs using planar electrodes,
coped with only by thin sensors and high bias voltages (see previous section).
Alternative pixel electrodes have been proposed and developed since the late 1990s \cite{3D-parker1997,DaVia:2013nn} as so-called 3D-silicon sensors featuring columnar electrode implants driven into the Si substrate perpendicular to the sensor surface (fig.~\ref{fig:3D-Si_sensors}). The smaller electrode distance than sensor thickness renders shorter drift distance and higher fields at moderate bias voltages, together resulting in an increased radiation tolerance.

Within the ATLAS IBL detector 3D-Si pixel sensors have been proven to operate well in a running experiment
\cite{IBL-paper}. After two years of operation the performance of 3D-Si pixel modules in terms of operation characteristics are on par with planar pixel modules, the latter operated with significantly higher bias \cite{LaRosa:2016nbd}. Optimizations for HL-LHC \cite{DallaBetta:2016_oct,Lange:2016jbm,DallaBetta:2016_review} include \cite{DallaBetta:2016_review} thinner sensors ($\sim$100\,$\upmu$m) on 6$''$ wafers, slimmer ($\sim$5\,$\upmu$m) and more closely spaced ($\sim$30\,$\upmu$m) electrodes, and very slim or active edges. A design example \cite{DallaBetta:2016_review} is shown in fig.~\ref{fig:3D-Si_new}.
In addition to cost and yield advantages of single-sided processing, studies have shown that the trade-off between signal efficiency and breakdown performance favors partial depth n-columns (not extending all the way through the thickness) \cite{dbetta2017}. The performance of such designs has been shown to yield high breakdown voltages before and after irradiation \cite{Sultan:2016vzg}.
The hit efficiency obtained with 3D-Si structures \cite{Lange:2017jrq} is demonstrated in
fig.~\ref{fig:3D-Si_performance}.

\begin{figure}
    \centering
    \subfigure[Efficiency map]{\raisebox{-0.2cm}
        {\includegraphics[width=0.53\textwidth]{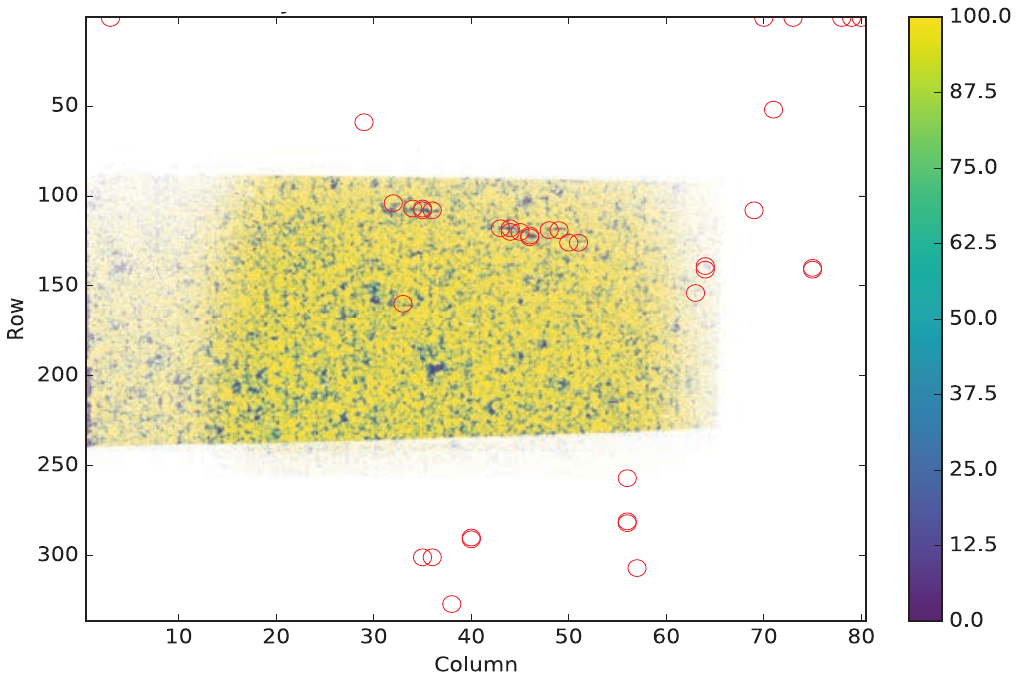}}\label{fig:DBM-efficiency}}\hskip 0.3cm
    \subfigure[3D-diamond assembly]{\raisebox{0.5cm}
        {\includegraphics[width=0.44\textwidth]{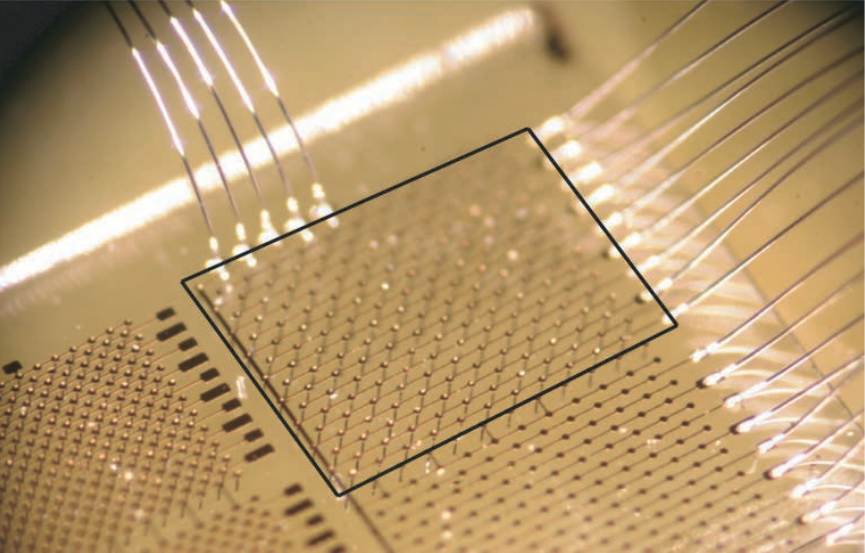}}\label{fig:3D-diamond}}
    \caption{Diamond pixels: (a) Efficiency map obtained in a high energy test beam showing the grain structure of the substrate. The mean overall efficiency is 87.6\%. (b) Assembled 3D device. \label{fig:diamond-pixels}}
\end{figure}

\paragraph{\bf Diamond pixels}
Diamond has been considered \emph{the} material for radiation hard pixels due to (a) its large band gap switching off any leakage current and (b) its twice as high energy kick-off threshold (43\,eV) to remove an atom from the lattice compared 25\,eV for Si, mitigating lattice damage at a given fluence. However, manufacturing and cost issues in producing single crystal sensor grade material (scCVD) as well as charge collection performance and other systematic issues associated with the grain structure of polycrystalline (pCVD) diamond, have so far prevented diamond pixels to
become real tracking devices in an experimental arrangement. However, large progress has been made in developing \glq
quasi tracker\grq\ like detectors, as for example the ATLAS Diamond Beam Monitor (DBM) consisting of four 3-layer telescopes arranged symmetrically around the beam at small forward and backward angles (see also \cite{Harris_HSTD2017}). An efficiency map \cite{Janssen:2017fko} exhibiting also the grain structure is shown in fig.~\ref{fig:DBM-efficiency}. With a mean overall efficiency of 87.6\% and remaining inhomogeneities due to the grain boundaries in pCVD, the demands on a stand alone tracking device for individual particles are not
yet sufficiently met. For applications, however, where robustness of response and radiation immunity
(e.g.\,for beam monitors) plays a larger role, pCVD diamond is a suitable choice.

The 3D technique can in diamond be realized by laser drilling sub-micrometer resistive holes \cite{Bani:2018ltd,Harris_HSTD2017}. A diamond 3D assembly is shown in Fig.~\ref{fig:3D-diamond}. With 150$\um^2$ cell size a signal of 13\,500 e$^-$ corresponding to a charge collection distance CCD > 350$\um$ was measured
with 92\% column efficiency \cite{Harris_HSTD2017}.

\begin{figure}
    \centering
    \subfigure[Hybrid pixels using CMOS sensors]{\raisebox{0.5cm}
        {\includegraphics[width=0.45\textwidth]{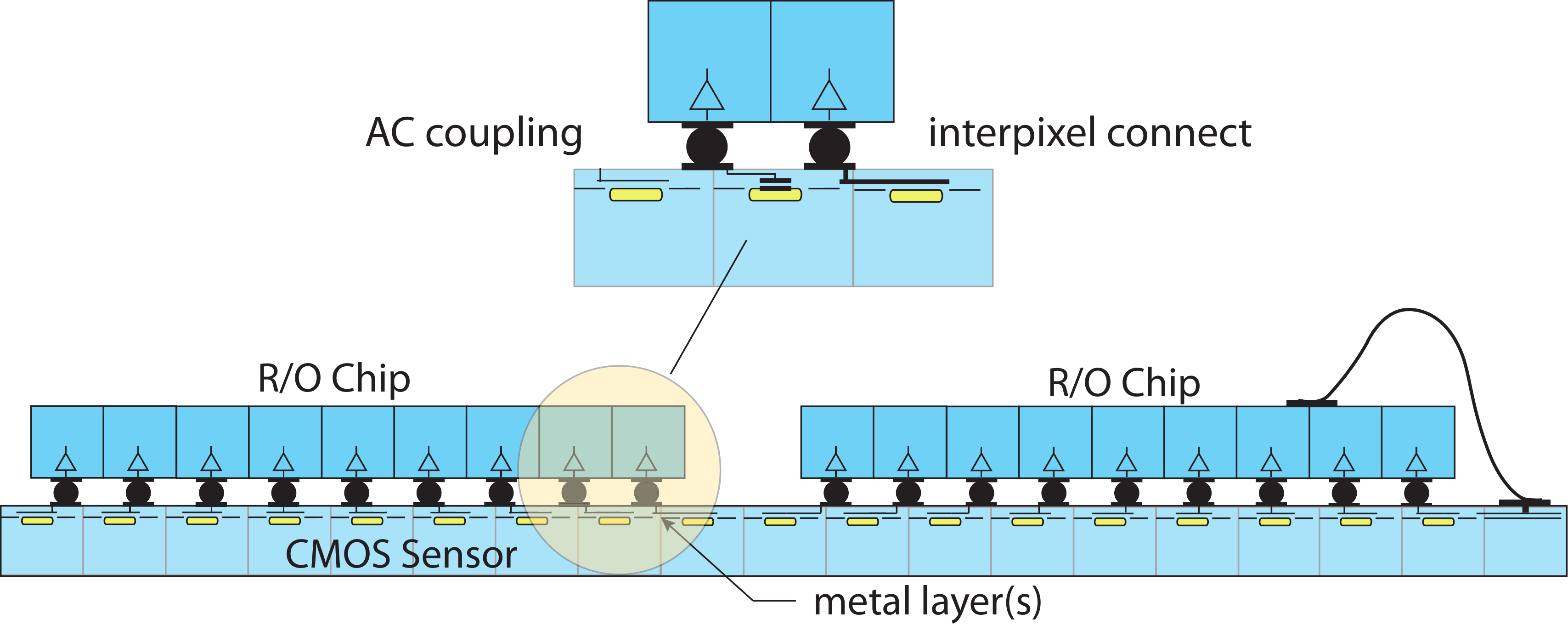}}\label{fig:Passive_CMOS_a}}\hskip 0.1cm
    \subfigure[Hit effciency]{\raisebox{0.0cm}
        {\includegraphics[width=0.45\textwidth]{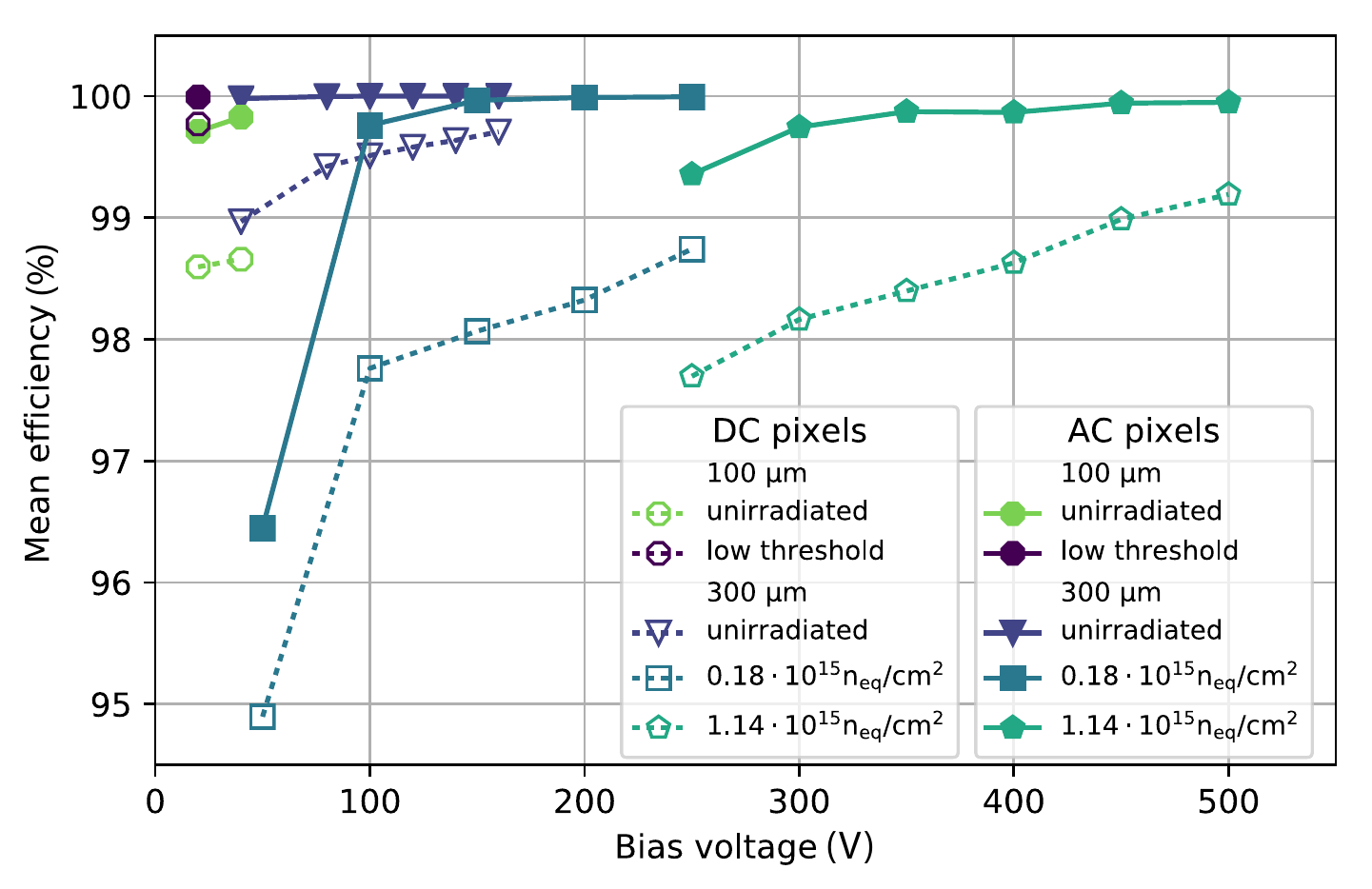}}\label{fig:Passive_CMOS_b}}
    \caption{Hybrid pixel module using passive CMOS pixel sensors (schematic). Depleted sensor employing CMOS technology with 1-2 metal planes that can be used for (i) AC coupling and (ii) rerouting. These features are detailed on top. The yellow area in every pixel denotes the charge collection node. (b) Hit efficiency measured in 3.2 GeV electron test beams~\cite{passCMOS_Bonn:2017} as a function of bias voltage, before irradiation as well as irradiated for two fluence levels.
    Close to 100\% efficiencies are achieves after a fluence of 1.1$\times 10^{15}$\,n$_{eq}$/cm$^2$ \label{fig:Passive_CMOS}.}
\end{figure}

\paragraph{\bf Sensors fabricated in CMOS production lines}
In the past high ohmic pixel sensors were fabricated by dedicated sensor production vendors. Employing high throughput chip wafer manufacturers for sensor fabrication was
believed to be too risky since the HEP community is not a large market to receive sufficient attention, and that the sensor quality would generally be insufficient. On the other hand, a number of advantages are in reach as there are: (i) Large volume production lines with price and turn-around benefits, (ii) 8$''$ or 12$''$ wafer sizes, (iii) wafers can be purchased to come with solder bumps of mid-size pitch ($150\um$), (iv) standardly available metal layers can be exploited for (a) AC coupling and (b) optimal line redistribution when connecting sensor to R/O chip, such that large and ganged pixels can be avoided (see fig.~\ref{fig:Passive_CMOS_a}).

The early conception is about to change as has been successfully demonstrated for strip \cite{Dragicevic:2014xja,Bergauer:2016uoq} and pixel sensors \cite{passCMOS_Bonn:2017}.
Measurements on passive CMOS pixel sensors, 100\,$\upmu$m and 300\,$\upmu$m thick, irradiated to fluences of 1.1$\times 10^{15}$\,n$_{eq}$/cm$^2$
have shown lab and test beam performance at least equal to those of planar sensors fabricated in dedicated sensor production lines \cite{passCMOS_Bonn:2017}. This is evident from fig.~\ref{fig:Passive_CMOS_b} where very high mean hit efficiency are reached for DC and particularly for AC coupled passive CMOS sensors bonded to the readout chip FE-I4 \cite{FE-I4}.

%
\section{Pixel readout chip}\label{sec:ROC}
Indeed, as said in the introduction, a large and complex readout chip with {\cal{o}}(10$^9$) transistors is a huge
and costly enterprise which has been addressed by the RD53 collaboration at CERN \cite{RD53}.
In comparison to the chips of the previous hybrid pixel generation which were column drain architectures without (1$^{\rm st}$ generation) or with (2$^{\rm nd}$ generation) a local (4-pixel) cluster efficient hit storage, this 3$^{\rm rd}$ generation contains architecture blocks with grouped logic enabling regional hit draining
surrounded by synthesized logic, dubbed the \glq digital sea\grq\ \cite{Demaria:2016dwz} and \cite{Maurice_HSTD2017}.

\begin{figure}
    \centering
    \subfigure[Layout example of the RD53A chip]{\raisebox{0.0cm}
        {\includegraphics[width=0.55\textwidth]{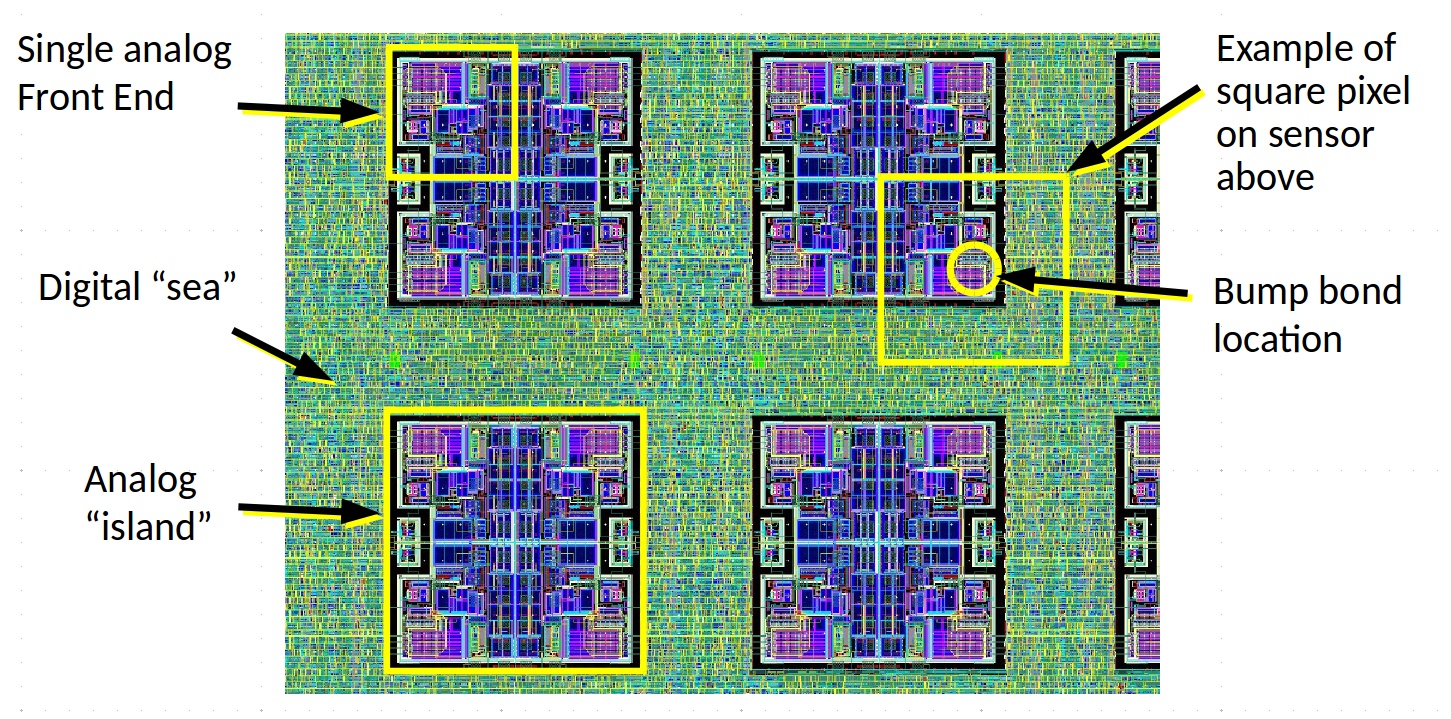}}\label{fig:RD53_layout}}\hskip 0.5cm
    \subfigure[Transistor cut]{\raisebox{0.0cm}
        {\includegraphics[width=0.35\textwidth]{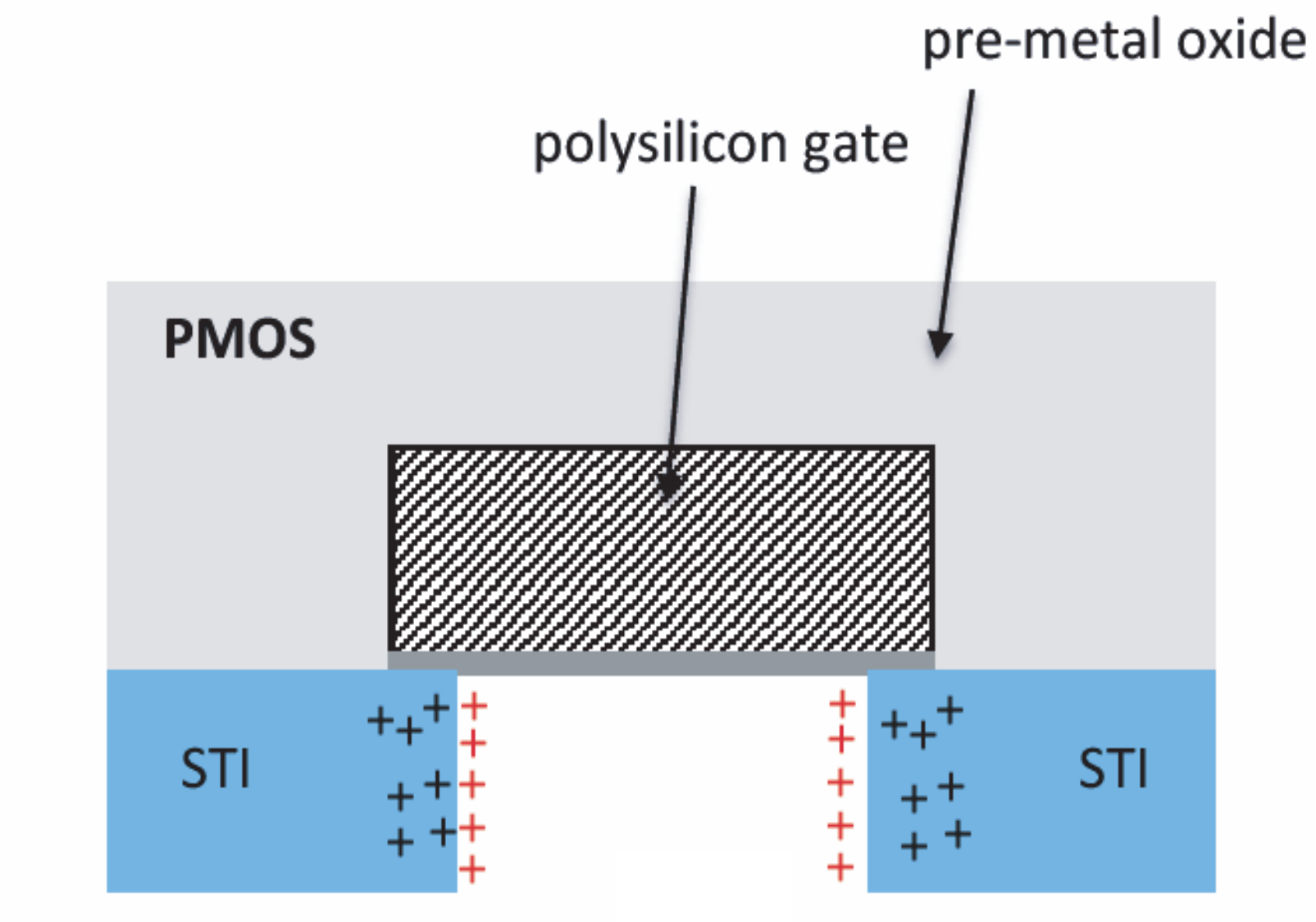}}\label{fig:RINCE}}
    \caption{Layout example of the RD53 chip \cite{RD53}. (a) Synthesized digital architecture surrounding analog regions (analog islands in digital sea); (b) Narrow and short channel effects cause new radiation vulnerabilities \label{fig:Passive_CMOS}.}
\end{figure}

Regarding radiation it seemed that the 250\,nm and 130\,nm technologies have hit a sweet spot. The naive belief
that smaller deep submicron technologies would be even more radhard neglected radiation induced narrow/short channel effects (RINCE and RISCE, respectively, Fig.~\ref{fig:RINCE}) as is detailed by F. Faccio's presentation at this conference \cite{Faccio:2015dc}.

%
\section{Monolithic Pixels}\label{DMAPS}
The many advantages of hybrid pixel detector face some serious drawbacks, among them the material budget and
the laborious assembly of the hybrid parts. In recent years monolithic pixel approaches have strongly emerged,
first by employing CMOS imaging technologies leading to so-called MAPS devices with charge collection
by diffusion in the chip's epitaxial layer \cite{Dierickx1998,MAPS-epi_Turchetta:2001}. This technology
has also found its way into HEP experiments in need for low material budget and small pixel devices, but
facing lower than LHC rates, such as RICH's STAR experiment \cite{RHIC_HSTD_2015,RHIC_Schambach_2015} and culminating in the 10\,m$^2$ pixel detector of the ALICE upgrade \cite{Abelevetal:2014dna} (see also \cite{mgsnw2018}).

It was believed for a long time that LHC rates and radiation levels were prohibitive for MAPS, but
they could be addressed by monolithic pixels in special processes, featuring
high voltage and high resistive wafers, multiple wells, and backside processing \cite{Peric:2007zz}.
In the past four years this approach culminated in fully monolithic designs that can cope with
the LHC radiation levels of above $10^{15}$n$_{eq}$cm$^{-2}$ (fluence) and 1\,MGy (dose) as well as the
corresponding particle rates. For the HL-LHC they are thus suited at least for the outer pixel layers.

\begin{figure}
\centering
   \subfigure[Large fill-factor]{\raisebox{0.0cm}
            {\includegraphics[width=0.45\linewidth]{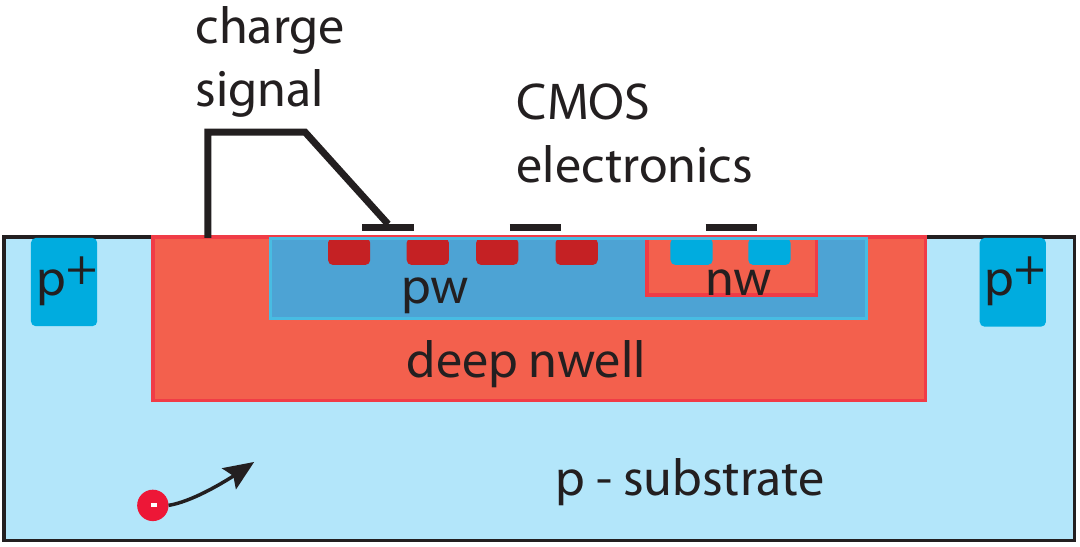}}\label{fig:large_FF}}\hskip 0.6cm
   \subfigure[Small fill-factor]{\raisebox{0.0cm}
            {\includegraphics[width=0.45\textwidth]{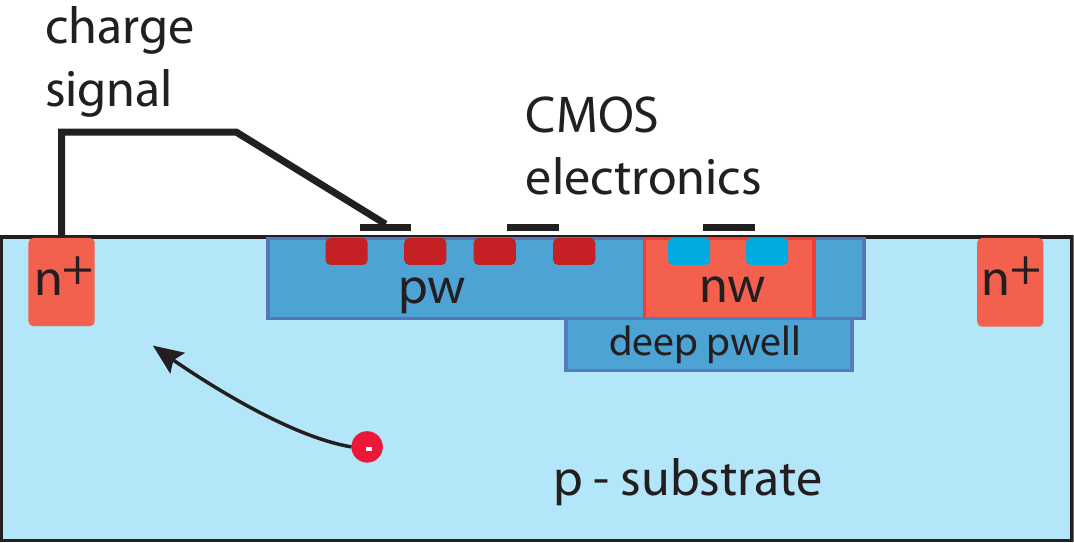}}\label{fig:small_FF}}
\caption{Two different CMOS cell geometries: (a) Large fill-factor: the charge collecting deep n-well encloses the complete CMOS electronics. (b) Small fill-factor: the charge collection node is placed outside the CMOS electronics area.}\label{fig:fill-factor}
\end{figure}
Two promising lines of actual research are currently followed: (a) large electrode designs (large fill factor, FF), where the electronics is embedded in the electrode covering a large fraction of the pixel area and (b) small electrode (small FF) designs where the collection electrode is set aside from the shielded electronics area (see also
\cite{Snoeys_HSTD2017}.)
The pros and cons are obvious. While the large FF design provides short average drift distances and hence promises
radiation hardness it suffers from a relatively large input capacitance to the amplifier, governed by its large area and, in addition, by the capacitance introduced between the deep p-well and the deep n-well. The latter also
represents a coupling path between electronics and sensor such that transient digital signals coupling
into the sensor must be prevented by dedicated circuitry. The small FF approach offers all the benefits imposed
by the small ($\lesssim 15$\,fF) input capacitance, suffers, however, from larger average drift distances for the
same pixel area. A process modification \cite{Snoeys:2017hjn} (see also \cite{Snoeys_HSTD2017}), strengthening the
lateral drift by sideward depletion, improves the charge collection towards the small collection electrode.

\begin{figure}
\centering
   \includegraphics[width=1\linewidth]{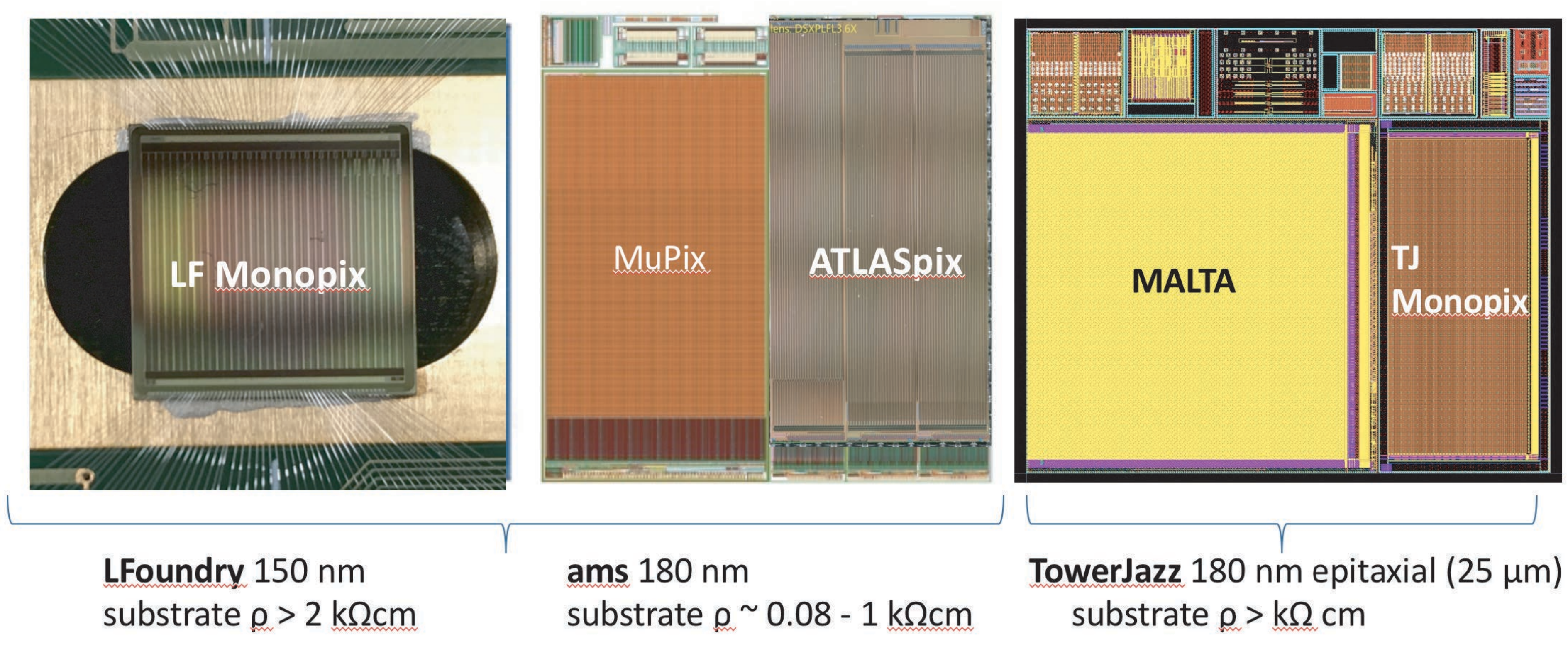}
   \caption{Fully monolithic depleted MAPS: (a) LF-Monopix in LFoundry 150\,nm technology containing a column drain and a \glq parallel pixel to bottom\grq\ (PPtB) read-out architecture; (b) \textsf{ams} aH18 180\,nm containing the MUPIX8 chip for the Mu3e experiment and several variants of the ATLASPix chip including PPtB read-out. (c) Small-FF design in TowerJazz technology: MALTA chip with asynchronous read-out, TJ-Monopix chip with column-drain read-out.}\label{fig:DMAPS-chips}
\end{figure}

Both approaches have been matured by intensive prototyping R{\&}D leading to large ($\sim$\,cm$^2$) fully monolithic pixel chips with full readout architectures suitable for LHC rates as shown in fig.~\ref{fig:DMAPS-chips}. The large FF chips are under test, the small FF design is currently fabricated (see also \cite{Pernegger_HSTD2017}).

\begin{figure}
\centering
   \subfigure[Depletion depth]{\raisebox{1.0cm}
            {\includegraphics[width=0.55\linewidth]{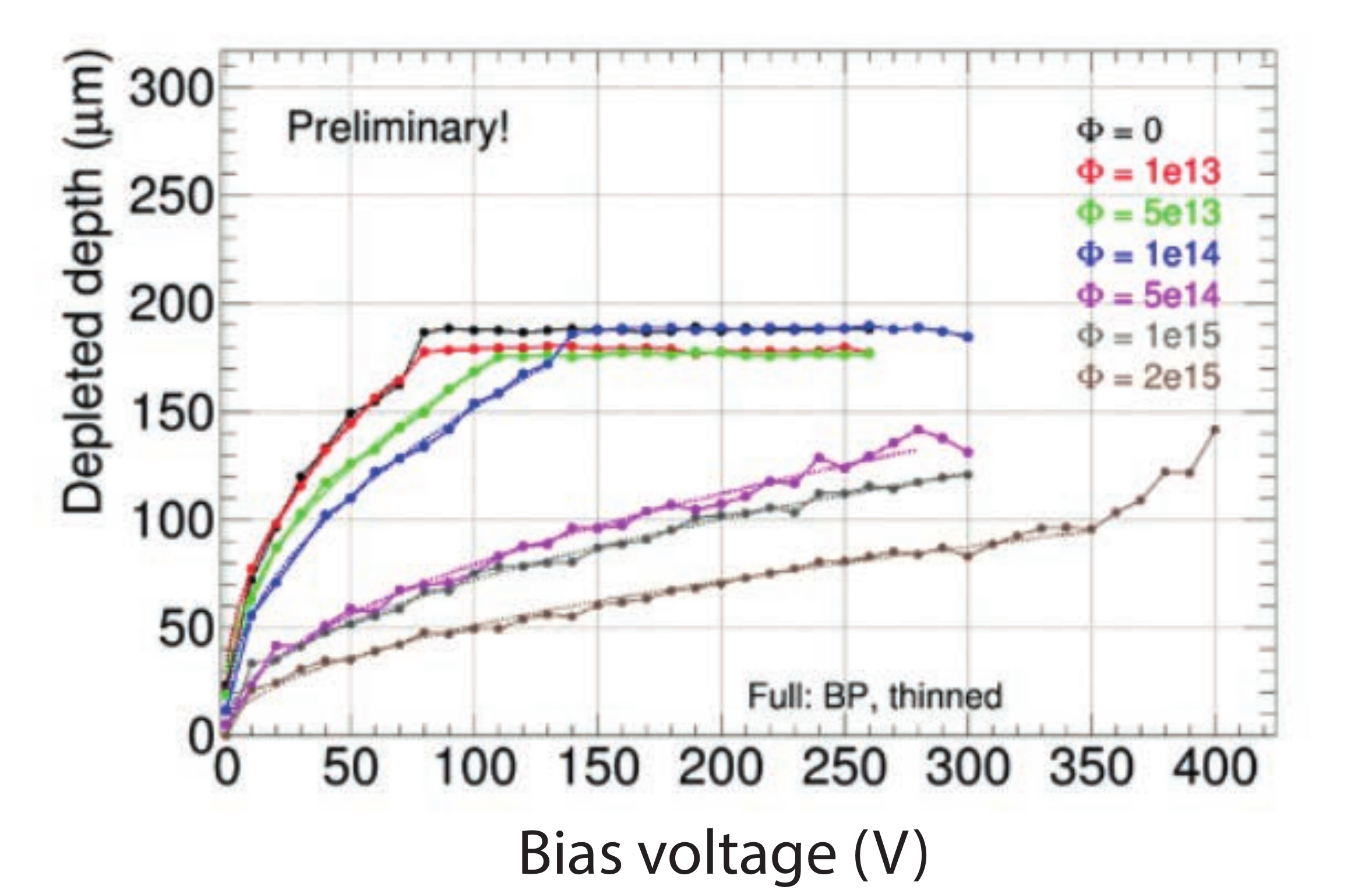}}\label{fig:LF_depletion}}\hskip 0.5cm
   \subfigure[Timing and Efficiency]{\raisebox{0.0cm}
            {\includegraphics[width=0.40\textwidth]{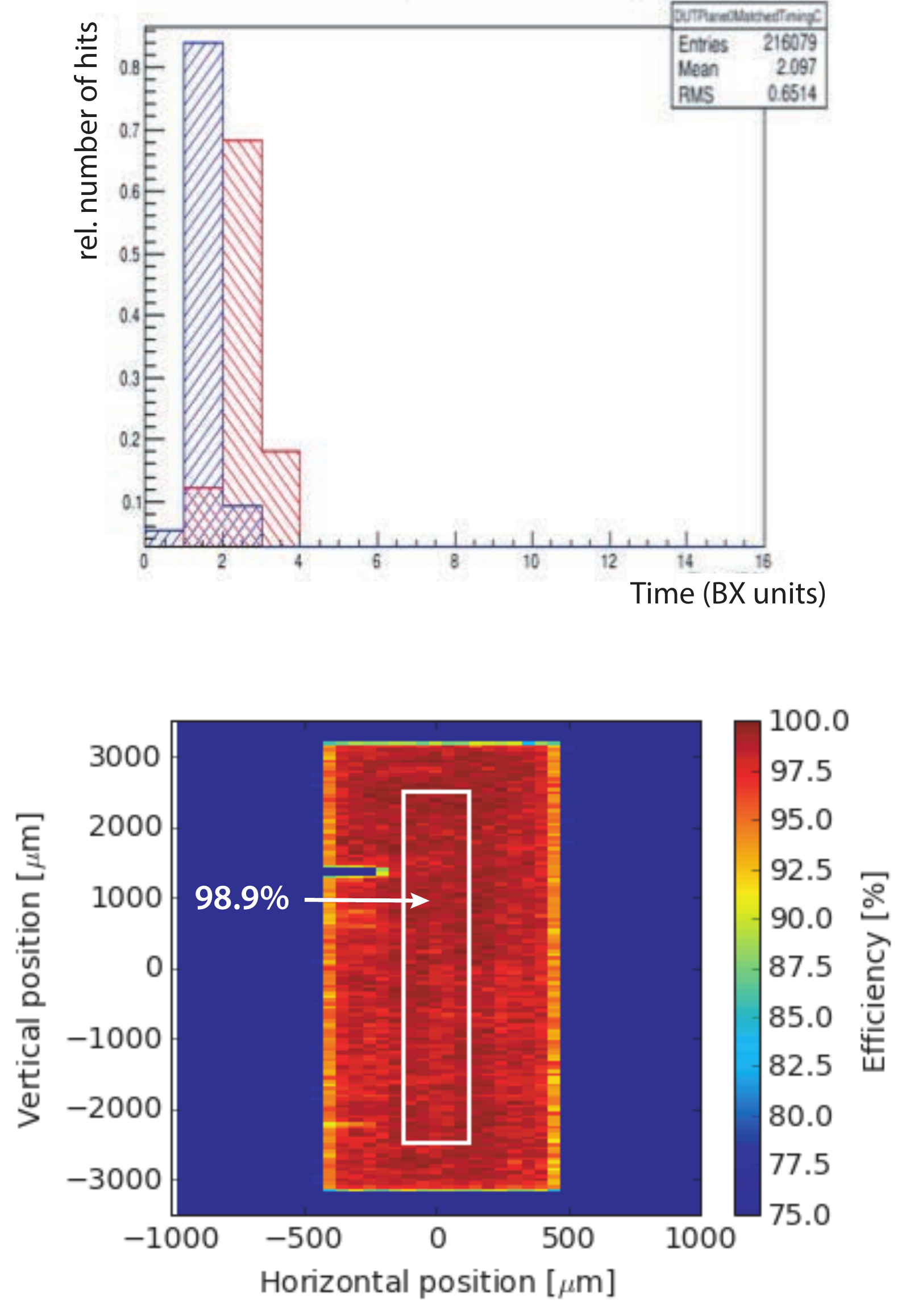}}\label{fig:ams_timing}}
\caption{Performance of depleted MAPS: (a) Measured depletion depth as a function of bias for various irradiation levels (LFoundry 150\,nm) \cite{Mandic:2018afg}. (b,\,top) Appearance of hits in bunch crossing time bins (ams 180\,nm); (b,\,bottom) hit efficiency (LFoundry 150\,nm), both after $10^{15}$n$_{eq}$cm$^{-2}$.\label{DMAPS_results}}
\end{figure}

The main results for large FF devices are summarized in fig.~\ref{DMAPS_results}: (i) Depletion depths in excess of 100$\um$ are reached even after $2\cdot 10^{15}$n$_{eq}$cm$^{-2}$ (fig.~\ref{fig:LF_depletion}); (ii) gain decreases and noise increases remain below 10\% and 30\%, respectively, after
radiation doses of 1\,MGy; (iii) the timing requirements after irradiation are close to being met after a correction; and (iv) the hit efficiency after $1\cdot 10^{15}$n$_{eq}$cm$^{-2}$ is still close to 99\% at a noise
occupancy of $10^{-7}$.

\paragraph{SOI pixels}
A full workshop within this conference is devoted to SOI pixels (see for example \cite{Arai_HSTD2017} and references therein). Mastering complex device variations in this so-called fully-depleted SOI technology the groups managed to
fabricate powerful detection devices for many application areas: particle tracking with sub-$\upmu$m resolutions (FPIX and SOFIST), X-ray imaging (INTPIX), X-ray astro (XRPIX, SOIPIX-PDD), synchrotron radiation (SOPHIAS), far infrared (cryogenic), biomedical counting (CNTPIX), and ion spectroscopy (MALPIX). A full account is given elsewhere at this conference.
Previously existing issues like the back-gate effect, coupling between circuit and sensor, and radiation (TID) issues
have been cured by several variants of buried well and nested well arrangements. The present devices are radiation
hard to TIDs of 100\,kGy.

Regarding SOI pixels for LHC radiation levels, it should be mentioned that promising prototyping has been done \cite{Hemperek:2014yoa} using partially depleted SOI, for which the back-gate and other effects do largely not exist due to shielding wells in the electronics layer.
%
\section{Fast timing with pixels}\label{sec:LGADS}
\begin{figure}
\centering
   \subfigure[LGAD structure]{\raisebox{0.0cm}
            {\includegraphics[width=0.60\linewidth]{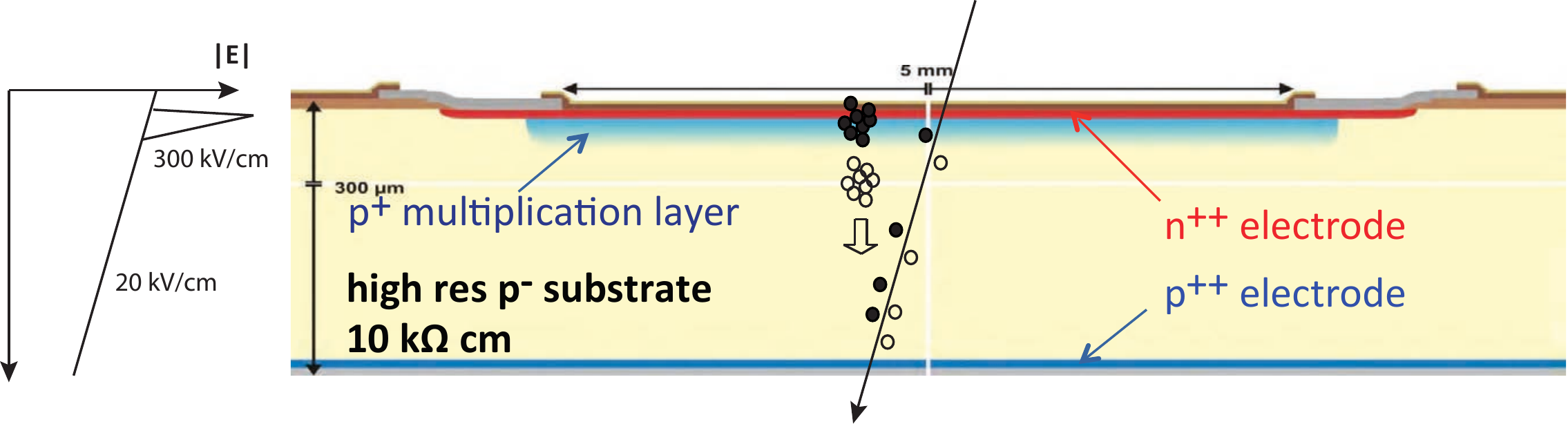}}\label{fig:LGAD_structure}}\hskip 0.5cm
   \subfigure[Achieved time resolutions]{\raisebox{0.0cm}
            {\includegraphics[width=0.35\textwidth]{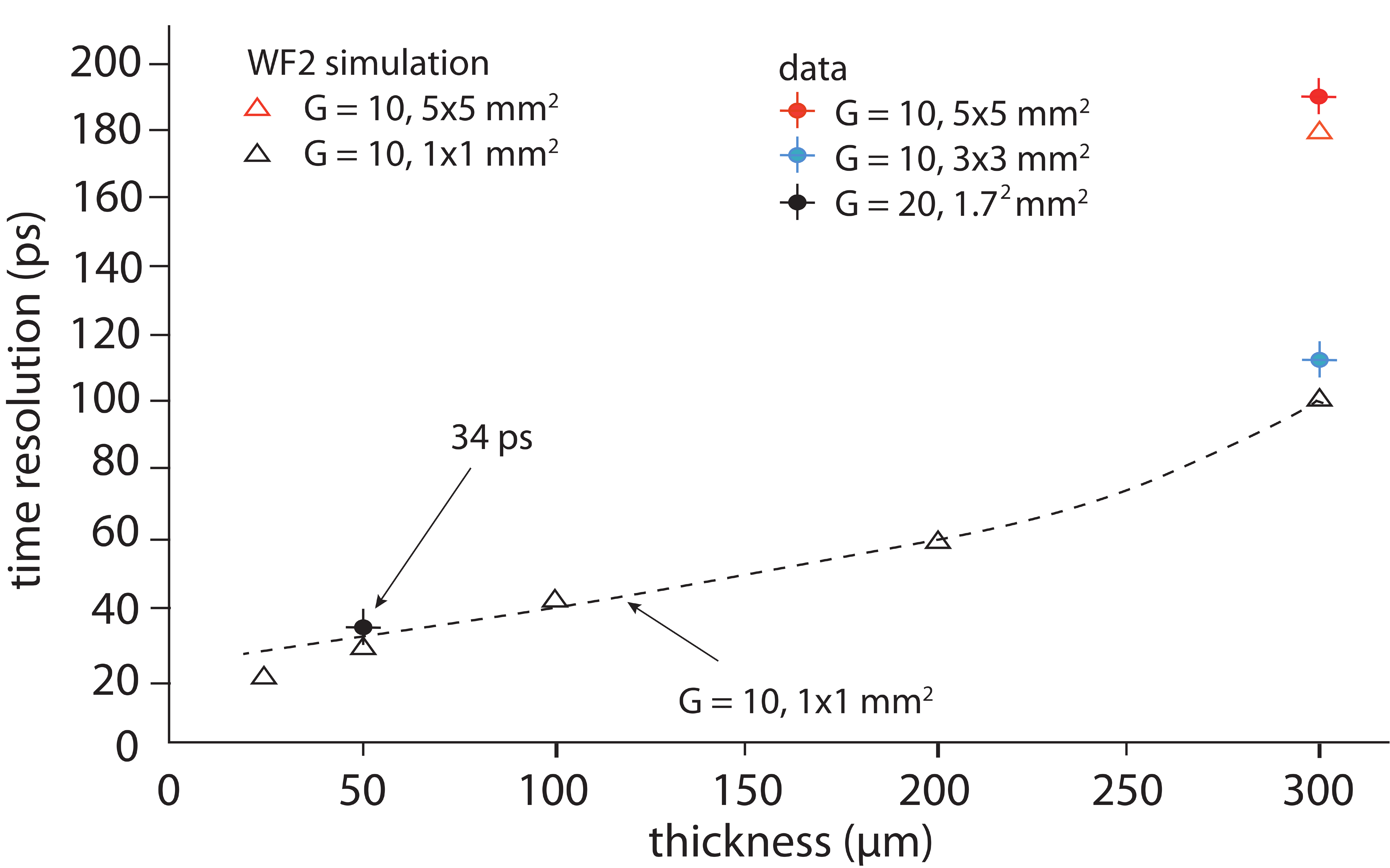}}\label{fig:LGAD_results}}
\caption{Low gain avalanche diodes are capable to provide high time resolutions. (a)
LGAD structure featuring a high ohmic p-type bulk, and an amplification junction (adapted from \cite{Cartiglia:2015iua}). The electric field is sketched on the left. (b) Comparison of time resolutions from simulations (WF2) and from test beam measurements. Data points from \cite{Cartiglia:2015iua},\cite{Sadrozinski:2016xxe},\cite{Cartiglia:2016voy},\cite{Cartiglia:2017yyy}, and \cite{Cartiglia:2016sjr}.
\label{LGADs}}
\end{figure}

Sub-ns to ps timing was believed to be very difficult with silicon detectors. So-called low gain avalanche diodes, having
mm$^2$ size patterns, have been developed to cope with this challenge (see also \cite{Sadrozinski_HSTD2017} and \cite{Sadrozinski:2017qpv}). In order to minimize time fluctuations in the signal generation process, an amplification structure is realized by a p$^+$ implantation right underneath the n$^{++}$ electrode (fig.~\ref{fig:LGAD_structure}). The fast signal is governed by fast e/h movements in high fields of thin detectors plus large slew rates ($dV/dt$) from amplified holes moving away from the readout electrode \cite{Sadrozinski:2013nja}.
Gains below 50 are targeted in order to avoid amplification excess noise. The structure needs a homogeneous weighting field such that small pixels are not easily realized.

The time resolution has several contributions \cite{Cartiglia:2016sjr}:
\begin{eqnarray}\label{eq:LGAD_time_resolution}
     \sigma_t^2 & = & \underbrace{\left( \frac{V_{th}}{dV/dt}\bigg|_{rms} \right)^2 } + \underbrace{\left( \frac{\rm Noise}{dV/dt}\right)^2} +\,\sigma^2_{\rm arrival} + \sigma^2_{\rm dist} + {\sigma^2_{\rm TDC} } \\
       && \hskip 0.5cm {\sigma_\textrm{time walk}^2}   \hskip 1.3cm {\sigma_\textrm{noise}^2}
\nonumber
\end{eqnarray}
The terms represent time walk, noise jitter, non-uniform charge depositions along a track, position dependent signal distortion, and digitisation. The first two contributions can be made small when large slew rates are achieved.

The results of this R{\&}D are encouraging as shown in fig.~\ref{fig:LGAD_results} comparing data and simulations
as a function of sensor thickness. In ATLAS and CMS proposals exist to employ LGADs for some timing applications for
the HL-LHC upgrade. The main concern currently is the radiation tolerance since the gain varies strongly with radiation fluence due to acceptor removal in the relatively highly doped p$^+$ layer of the multiplication region.
%
\section{Imaging with Hybrid Pixels}\label{sec:imaging}
As mentioned in the introduction, the hybrid pixel developments made for particle tracking shortly afterwards
entered X-ray imaging for biomedical (e.g. MEDIPIX \cite{Ballabriga:2017yrw}) or synchrotron light applications
(see also \cite{Carini_HSTD2017}). The first big detectors addressing
signals with huge dynamic range (1 - 10$^5$) have come into live. Examples are among others the series of developments at PSI called EIGER (500\,k pixels, 75\,$\upmu$m pitch, photon counting, 23 kHz frame rates), M\"ONCH (charge integrating, 25$\um$ pixel pitch, low noise, low energies) \cite{Ramilli:2017ojx}, and JUNGFRAU (charge integrating, 75$\um$ pixel pitch, dynamic gain switching) \cite{Mozzanica:2016vro} for the SLS and for SWISSFEL, with which a step function in synchrotron light imaging is achieved. For the EUROPEAN XFEL first pictures of the
adaptive gain imaging pixel detector AGIPD \cite{Allahgholi:2016svx} have been recorded.

Also in imaging applications monolithic pixels are now taking an important role, as can for example by seen be the
\glq Double SOI\grq\ pixel detector presented at this conference \cite{Miyoshi_HSTD2017}.
\section{Conclusions}
Pixel detectors have paved the way of high resolution, high rate, and high radiation devices indispensable from
particle tracking and imaging experiments. The path first laid by hybrid pixel detectors is now followed into
monolithic devices and devices tuned for high timing resolution, at first again for tracking.
For sure imaging applications will follow.

\section*{Acknowledgments}
The author would like to thank the organizers for their kind invitation.
This work was supported in parts by the Deutsche Forschungsgemeinschaft DFG,
grant number WE 976/4-1, by the German Ministry BMBF under
grant number 05H15PDCA9, and by the H2020 project AIDA-2020, GA no. 654168, and by the H2020
project STREAM, GA no. 675587.

\begingroup
\raggedright 
   \bibliographystyle{unsrt}
    \bibliography{Wermes_HSTD2017}
\endgroup

\end{document}